\documentclass{sigchi}

\toappear{
Article currently under peer review for publication
}

\usepackage{balance}  
\usepackage{graphics} 
\usepackage{times}    
\usepackage{url}      

\usepackage[utf8]{inputenc}  

\makeatletter
\def\url@leostyle{%
  \@ifundefined{selectfont}{\def\UrlFont{\sf}}{\def\UrlFont{\small\bf\ttfamily}}}
\makeatother
\def\pprw{8.5in}
\def\pprh{11in}

\usepackage[pdftex]{hyperref}
\hypersetup{
pdftitle={Evaluating arbitration and conflict resolution mechanisms in the Spanish Wikipedia},
pdfauthor={LaTeX},
pdfkeywords={Wikipedia, Arbitration, Conflict resolution, CRC, Arbcom},
bookmarksnumbered,
pdfstartview={FitH},
colorlinks,
citecolor=black,
filecolor=black,
linkcolor=black,
urlcolor=black,
breaklinks=true,
}

\newcommand\tabhead[1]{\small\textbf{#1}}

\begin{document}

\title{Evaluating arbitration and conflict resolution mechanisms in the Spanish Wikipedia}

\numberofauthors{2}
\author{
  \alignauthor Maria Sefidari\\
    \affaddr{University Rey Juan Carlos}\\
    \affaddr{Tulipan s/n, Mostoles. Spain.}\\
    \email{mk.sefidari@alumnos.urjc.es}\\
  \alignauthor Felipe Ortega\\
    \affaddr{DEIO, University Rey Juan Carlos}\\
    \affaddr{Tulipan s/n, Mostoles. Spain.}\\
    \email{felipe.ortega@urjc.es}\\  
}

\maketitle

\begin{abstract}
In open collaborative projects like Wikipedia, interactions among users can produce tension and
misunderstandings. Complex disputes require more sophisticated mechanisms of conflict resolution.
In this paper, we examine the case of the Spanish Wikipedia and its Arbitration Committee, 
known as CRC, over its two years of activity. We postulate that the high percentage of 
rejections of cases presented by non-administrators, the lack of diversity inside the 
committee (composed only by administrators), and the high number of cases involving administrators 
played a central role in its eventual downfall. We conclude that mechanisms that fail to
acknowledge the ecosystem they are part of cannot succceed. Therefore, further research is 
needed to determine if granting more decision-making power to non-administrators may lead 
to more effective conflict resolution mechanisms.
\end{abstract}


\category{H.5.3.}{Group and Organization Interfaces}{Web-based interaction}

\section{Introduction}

Wikipedia is an online free encyclopedia that anyone can edit. You only require a 
computer (or tablet, or smartphone) and access to an Internet connection to make an 
edit in an article of your choice. You don’t even need to create an account to do so: 
more than half of all total contributions to Wikipedia come from anonymous users, identified 
only by their IP (Internet Protocol) address~\cite{rfc791}. People are encouraged to edit Wikipedia, 
and one of its main guidelines is Be Bold! when contributing to articles.

However, its very open and collaborative nature can produce tensions unlike in traditional
encyclopedia-creation processes. A common example occurs when a user adds information, and 
someone else corrects or removes it. Wikipedia policies highlight it is not a social site, 
but social interactions do happen at all times in it. And it is those interactions that can
produce conflicts and misunderstandings which, if unaddressed, can impact the quality of the
encyclopedia. Wikipedia’s strategy is completely based in the use of volunteers: having an
unhealthy working environment, ongoing conflicts and fights which drain the time and mental 
health of contributors can lead to volunteer burnout and abandonment of the project. So how 
to deal with conflict disputes and how to adequately resolve them is of the utmost importance 
~\cite{kittur2010}. Examining what happened with the failed conflict resolution system in a project like 
Spanish Wikipedia could provide insights into what kind of pitfalls other similar projects 
could and should avoid in order to have effective dispute resolution mechanisms.

Both English and Spanish Wikipedia share the same pillar, Etiquette~\cite{wiki:wikidata}, which encourages 
users to be civil to each other as everyone shares a common purpose of improving content. 
Other policies and guidelines regulate what is allowed and what is not when addressing other
fellow contributors\cite{butler2008, morgan2010}. Insults and attacks are not 
allowed, and can be fair cause to be
blocked from the project. Some policies regulate the behavior when adding or removing content, 
and infractions can result in warnings and blocks. When a user believes another user has broken 
a rule or policy, they can report it to an administrator, a trusted user who has the technical
ability to delete or protect pages, and to block users among other things~\cite{wiki:userlevels}. 

However, when a case is too complex for one administrator to deal with, the case will be 
referred to a different
setting. This could be Mediation, the Administrators’ Noticeboard, the Village Pump, or the
Arbitration Committee. This last instance is the focus of our current study. The English 
Wikipedia has had a functional Arbitration Committee for several years, that is still operational. 
Instead, the Spanish Wikipedia had a similar instance for conflict arbitration running for two 
years, until the community of editors resolved to take it down. In this paper, our goal is to
analyze public data describing the different arbitration requests elevated to this committee to
identify possible reasons that may have led to its closure. In the following section, we present
the data obtained for our study and the approach followed to carry out the analysis. Then,
Section "Results" presents the main outcomes from the study, whereas Section "Discussion" evaluates
the main inferences that can be drawn from results regarding the effectiveness of this arbitration
committee. Finally, the last section concludes the paper summarizing the main lessons for better
design of conflict resolution mechanisms and outlines further lines of research on this topic.

\section{Methodology}

In 2003 the English Wikipedia created the Arbitration Committee (ArbCom, for short), as a result 
of the huge growth and success experimented by the project~\cite{wiki:arbcom, riehle2006}. 
Wikipedia co-founder Jimmy Wales could no longer cope with an increasing number
of arbitration requests in dispute resolutions.
As a result of this, a community-elected committee took in charge of examining and resolving
complex cases that could not be easily decided and which required painstacking examination of
contributions over time, as well as allegations from involved and non-involved users. 

Later on, the Spanish Wikipedia
replicated this model and created its own Arbcom, named \textit{Comité de Resolución de Conflictos}
(CRC, for short), on December 12, 2006. The first committee was community-elected on January 15, 2007, and was composed only of admins. During its two years of activity (2007 and 2008), no regular user
without the administrator level was ever integrated in this committee. The CRC was composed of seven active members plus two backup members.

Though internal deliberations of the CRC committee were secret (via a private mailing list and wiki),
dismissals and resolutions were archived and published for the benefit and transparency in the community~\cite{wiki:CRC-archivo}. In
consequence, all the cases (89) presented to the CRC during 2007 (58) and 2008 (31) for their
consideration were openly published on Spanish Wikipedia, along with the global rate of accepted
cases and dismissals. In this study, we manually retrieved
and analyzed publicly available data from the wiki pages describing these 89 cases. User data 
such as date of first contribution, number of edits or whether a user involved in the case is an
administrator or not is also available in the public summary for each case. Then, the analysis of 
every case follows, exploring different useful metrics to assess the effectiveness of this conflict
resolution mechanism, including: 

\begin{itemize}
	\item the user (login) name of editors involved in the case;
    \item for how long the user who opened a case had been contributing to Wikipedia;
    \item whether the person who opened the case is an admin or not;
    \item whether there are admins involved in the case or not;
    \item if the case is rejected or granted;
    \item if the user who opened the case receives a warning or not;
    \item how long it took the committee to resolve the case.
\end{itemize}

Based on this information, we perform an observational study using these data
to build a synthetic \textit{postmortem review}~\cite{Dingsoyr:postmortem}. From
this analysis, we assess the performance of the CRC in the Spanish Wikipedia
as a conflict resolution mechanism, possible factors leading to its ending and
important lessons for the design of similar arbitration mechanisms in open
collaborative communities.

\section{Results}

\subsection{Types of users}

We distinguish here between type of users who open a case, and types of users involved in the
conflict as named parties. Before opening a case, the average user had been contributing to
Spanish Wikipedia for 16.7 months. The average for 2007 is 13.5 months, and the average for 2008
is 22.6 months. 89.1\% were non-admins in 2007 and 66.7\% were non-admins in 2008. Total average
of non-admins opening a case was 77.9\%.
As for the type of users involved with the cases, an administrator was involved as one of the
named parties of the conflict in 78\% of total cases (79\% for 2007 and 74\% for 2008).

\begin{table}
  \centering
  \begin{tabular}{|c|c|}
    \hline
    \multicolumn{1}{|p{0.15\columnwidth}|}{\centering\tabhead{Year}} &
    \multicolumn{1}{|p{0.2\columnwidth}|}{\centering\tabhead{Admins involved}} \\
    \hline
    2007 & 79\% \\
    \hline
    2008 & 74\% \\
    \hline
    Total & 78\% \\
    \hline
  \end{tabular}
  \caption{Administrators involved as named parties of the conflict by year.}
  \label{tab:table1}
\end{table}

\subsection{Dismissals and resolutions}
When a case was presented to the CRC, it could be accepted and go forward to achieve a resolution,
which could be favorable or negative for the user who opened the case, or it could instead be
dismissed, which meant the CRC would not take the case. 90\% (62 out of 69) of total cases
presented by non-admins were dismissed by the CRC. 88\% were dismissed in 2007, and 95\% were
dismissed in 2008. If it was an admin presenting a case, it was accepted 44\% of the time (67\% 
in 2007 and 30\% in 2008). 

\begin{table}
  \centering
  \begin{tabular}{|c|c|}
    \hline
    \multicolumn{1}{|p{0.15\columnwidth}|}{\centering\tabhead{Year}} &
    \multicolumn{1}{|p{0.2\columnwidth}|}{\centering\tabhead{Admins involved}} \\
    \hline
    2007 & 88\% \\
    \hline
    2008 & 95\% \\
    \hline
    Total & 90\% \\
    \hline
  \end{tabular}
  \caption{Cases presented by NON-ADMINS and dismissal rate by year}
  \label{tab:table2}
\end{table}

Of those cases presented by non-admins and accepted by the CRC, 22\% would have a resolution 
that included a warning to the user who opened the case to not misuse the CRC by filing cases 
or face sanctions. 75\% of all accepted cases ended with an unfavorable resolution for the
claimant. Breaking it down, in 2007 63.6\% of cases ended with an unfavorable resolution for the
claimant, while in 2008 86.7\% of cases ended with an unfavorable resolution. Only 25\% of all
accepted cases both years were favorable to the claimant: 36.4\% in 2007 and 13.3\% in 2008. 
These include requests to become a Checkuser and one investigation initiated by the CRC itself.

\begin{table}
  \centering
  \begin{tabular}{|c|c|}
    \hline
    \multicolumn{1}{|p{0.15\columnwidth}|}{\centering\tabhead{Year}} &
    \multicolumn{1}{|p{0.2\columnwidth}|}{\centering\tabhead{Admins involved}} \\
    \hline
    2007 & 67\% \\
    \hline
    2008 & 30\% \\
    \hline
    Total & 44\% \\
    \hline
  \end{tabular}
  \caption{Cases presented by ADMINS and acceptance rate per year}
  \label{tab:table3}
\end{table}

\begin{table}
  \centering
  \begin{tabular}{|c|c|c|}
    \hline
    \multicolumn{1}{|p{0.15\columnwidth}|}{\centering\tabhead{Year}} &
    \multicolumn{1}{|p{0.25\columnwidth}|}{\centering\tabhead{Favorable resolution}} &
    \multicolumn{1}{|p{0.25\columnwidth}|}{\centering\tabhead{Unfavorable resolution}} \\
    \hline
    2007 & 36.4\% & 63.6\%\\
    \hline
    2008 & 13.3\% & 86.7\%\\
    \hline
    Total & 25\% & 75\%\\
    \hline
  \end{tabular}
  \caption{Favorable vs unfavorable resolution for the claimant per year}
  \label{tab:table4}
\end{table}

\subsection{Time}
The time between the filing of a case and the CRC not accepting it or publishing a resolution 
was also measured. The average days it would take the CRC to close a case in 2007 was 36.4. 
The median was 16 and the mode zero. In 2008, the average was of 55.8 days, the median was 34 
and the mode 99. The total average for the CRC to close a case was 43.9 days, median 23.5, 
mode zero. In 2007 there were 8 cases that had no acceptance or resolution, either because 
the case was archived by the CRC or the filer withdrew it. In 2008, there was only one.

\section{Discussion}

The CRC was dissolved after a series of community debates about its efficiency and after a 
final community election that ended on April 2009~\cite{wiki:disolCRC}. But before that, during December 2008, 
the community had already voted to suspend the committee for the first semester of 2009 while 
they decided what to do with it, so no new members were appointed and no new cases were taken 
by the CRC starting January 1st, 2009~\cite{wiki:suspenCRC}.
The CRC was regarded as the last option to solve a conflict. There were four possible motivations
under which anyone could file a case and the CRC would act: a disagreement over content in an
article after a series of steps had taken place, a disagreement over the actions of an admin who
had allegedly abused their tools, a disagreement over the actions of users who had violated
policies, and lastly, there was the possibility that something could alter the normal functioning
of the project, and so the CRC had the prerogative to initiate a case of its own 
accord~\cite{wiki:motivacion}.

The CRC could also grant a user the \textit{Checkuser} tools (\url{http://en.wikipedia.org/wiki/Wikipedia:CheckUser}) as per global policy. 
What we see in our results is that non-admins lean to initiate cases that involved one or
several administrators as named parties of the conflict. The CRC, composed only by admins during
its entire period of activity, either did not accept or decided to reject a high number of these.
90\% of total cases presented by non-admins were dismissed by the CRC. 22\% of accepted cases
ended with warnings to the non-admin to not misuse the CRC. Only 25\% of accepted cases ended up
having a favorable resolution for the claimant. 

There are two possible interpretations for this:
either the majority of the claimants had a different view of what the role of the CRC was, or the
CRC had a different view of what its purpose was. A committee tasked with solving conflicts which
dismisses the majority of cases presented before it is probably generating more conflict than it
solves. At the very least, it shows there was a dissonance between the committee and the community
that wished to use it. 

The community did recognize this and attempts to reform the CRC were made.
But this did not happen and eventually the committee was dissolved and unmissed. Some factors that
may help explain this was the difficulty to explain what it was that did not work. Several
committee members did attempt self-criticism, particularly regarding the time it took for the
committee to deal with open cases, and the need to change the required motivations to open a case
so they could be more flexible and accept more. But while that may have been a factor in the
dissatisfaction of the community, there were probably others. 

For instance, one such factor could
be a particular lack of diversity in the composition of the CRC. At the time when the CRC existed,
there were less than 100 active administrators in the project. The rest of the users were
non-admins. However, the committee never had non-admins, arguably the biggest constituency of the
project, in it. One possible explanation is that being an admin is automatically equated with
being trustworthy, with having the support of the community to perform extra tasks. In an election
for CRC members it would be a plus to be an admin, and a committee composed only by admins would
be a very trustworthy committee. 

Nonetheless, the lack of different points of view from a user perspective
could have been very detrimental to the performance of the committee. Administrators were involved
as named parties of the conflict in 78\% of the total cases presented to the CRC. It is possible a
committee composed not only by admins would have been more likely to not dismiss 90\% of all total
cases presented by non-admins. This would have required the community at large to acknowledge, not
only in theory but in practice as well, that non-admins can hold decision-making positions in the
community and be just as trustworthy as admins are required to be. If the committee had accepted
more cases and had been perceived as more useful to the community, it would have complied with its
community-mandated mission to help reduce conflicts. 

In the end, the community perceived that
conflicts successfully resolved by the CRC were not enough to justify its existence, and decided to
do away with it. The takeaway lesson here would be that in the context of an ecosystem 
with different types
of users, if only one type has decision-making capabilities, those decisions are hardly going to
be a reflection of the diversity or the needs of said ecosystem. And when that happens, the
ecosystem (the “community”) will take it down. While ArbCom in itself can be in theory a useful
tool to resolve conflicts, the system can crash if it fails to acknowledge the ecosystem in which
it works, and the needs of its community.

\section{Conclusions and further research}

English Wikipedia ArbCom requires identification to the Wikimedia Foundation and there is
oversight. Spanish Wikipedia never evolved to this point before it was dissolved. It would be
interesting to see if these factors play a part in holding the committee accountable to the wider
community.
While not all projects have an ArbCom, several have an Administrators’ Noticeboard, including
Spanish Wikipedia. These usually are the last instance of conflict resolution. In these
administrators’ noticeboards, the decision-making also is usually in the hands of admins only,
although some projects allow the comments of non-admins before an admin resolves the issue. Is
this tenable in the long-term? Could the lack of diversity be a negative factor in the conflict
resolution in those projects, leading to difficulties to retain its contributors? Could this also
have an effect on the Requests for Adminship, if the role of administrators has moved from a
merely technical oversight to decision-making? Future research could help
elucidate the questions to these answers. 

\section{Acknowledgments}

We thank L. Hale for her technical support to undertake this study.

%
%
%
%
%
\balance


\bibliographystyle{acm-sigchi}
\bibliography{sample}
\end{document}